# Laser-Ablated Au Electrodes with Preferred Orientation and Flat Interface on Oxides

Ambrose Seo[a,*], Richard Lai[a], Sujan Shrestha[a], Tina Tong[a], Joseph W. Brill[a], Menglin Zhu[b], Chris Chae[b], and Jinwoo Hwang[b]

[a] *Department of Physics and Astronomy, University of Kentucky, Lexington, KY 40506, USA*
[b] *Department of Materials Science and Engineering, The Ohio State University, Columbus, Ohio 43210, USA*

**Abstract:**

Gold (Au) thin-film electrodes deposited on oxides by pulsed laser deposition (PLD) exhibit distinctive structural properties compared with thermally evaporated Au films. While thermally evaporated Au forms polycrystalline films with randomly oriented grains, PLD-grown Au films show a strong preferred [111] orientation perpendicular to the substrate surface, likely due to the energetic nature of the laser-ablated Au plume. High-resolution transmission electron microscopy reveals atomically flat interfaces between the PLD-grown Au electrodes and epitaxial oxide thin films, separated by nanometer-scale gaps comparable to interlayer spacings in van der Waals materials. The highly ordered microstructure and flat interfacial morphology result in enhanced adhesion of the Au films to oxide surfaces. These results demonstrate that PLD can produce both high-quality Au electrodes and oxide thin films within a single deposition platform, offering improved interfacial control and potential benefits for a wide range of oxide electronic and functional devices.

Correspondence to: a.seo@uky.edu

## 1. Introduction

Gold (Au) stands as the most extensively employed electrode material in modern electronics, due to its high conductivity and corrosion-resistant properties. These characteristics make it an essential component in electronic applications such as complementary metal-oxide semiconductor (CMOS) technology, ensuring unwavering performance over extended timeframes. Generally, the deposition of polycrystalline Au electrodes onto patterned surfaces of electronic devices is accomplished through physical vapor deposition methods like thermal evaporation and sputtering. However, polycrystalline Au electrodes are known for their uneven interfacial formation and diminished adhesion to oxide or semiconductor surfaces. To enhance the interfacial characteristics of Au electrodes in devices, additional measures are often necessitated in the device fabrication process. These include the deposition of intermediate layers such as titanium or chromium [1-3], as well as various surface preparation techniques like ion bombardment [4] and plasma treatment [5, 6], which involve etching or roughening the surface. While these steps are successfully applied in CMOS technology, they simultaneously escalate costs and can substantially impact device performance [7-9].

We have noted that Au films deposited through sputtering exhibit good surface adhesiveness in comparison to thermal evaporation [10], regardless of the substrate's surface roughness. This is likely due to the heightened energy of the Au vapor during sputtering. A pulsed laser deposition (PLD) technique is renowned for its highly energetic plume [11], making it a compelling avenue for further improving the Au electrode's interfacial properties. Since PLD is widely used for synthesizing epitaxial oxide thin films, it would also enable sequential depositions of both oxide layers and Au electrodes in a single deposition platform. Therefore, we have investigated the impact of PLD on the properties of Au electrodes and their interfacial properties with oxide thin films.

In this paper, we present our findings on Au film electrodes deposited onto oxide surfaces using PLD, revealing distinct structural and interfacial characteristics. The Au electrodes grown by PLD on oxides exhibit a notable [111] crystallographic orientation along the surface normal, in contrast to thermally evaporated Au films that have polycrystalline grains with random orientations. Remarkably, high-resolution scanning transmission electron microscopy indicates the formation of a nanometer-scale flat gap at the interface between Au electrodes and oxide thin films. This observation suggests that heightened interfacial interactions might have been formed between coherently oriented Au (111) planes and epitaxial oxide surfaces. We propose that the generation of highly energetic Au plumes through pulsed laser ablation is noteworthy in the development of desired Au electrodes, imparting advantages to electronic device fabrication processes.

## 2. Experimental methods

We performed the deposition of Au films on a range of oxide surfaces utilizing both PLD and thermal evaporation techniques. The surfaces subjected to investigation encompassed commercially available glass plates (i.e., amorphous silica), single crystal oxide substrates such as $SrTiO_3$ (STO), and epitaxial oxide thin films. Prior to Au film deposition, the glass plates underwent thorough cleaning using typical solvents such as acetone and methanol. Atomically flat surfaces of the STO substrates were prepared through thermal annealing and subsequent deionized water leaching, as outlined in Ref. [12]. All depositions, involving approximately 50-nm-thick Au films, were carried out at room temperature within a vacuum chamber maintained at a pressure of around $10^{-7}$ Torr. For PLD, we employed an ultraviolet excimer laser operating at a wavelength of 248 nm, with an energy density of 3 J/cm$^2$ and a pulse duration of 20 ns. The Au target and wires (purity of 99.99%) were commercially sourced.

## 3. Results and discussions

### *3.1 X-ray diffraction of Au films grown by pulsed laser deposition and thermal evaporation*

The PLD-grown Au films show a preferential orientation [111]-orientation along the surface normal direction, whereas thermally evaporated Au films are randomly oriented polycrystals. Figure 1 shows X-ray diffraction (XRD) $\theta$-$2\theta$ scans of the PLD-grown Au films featuring only two XRD peaks, i.e., (111) and (222) planes, on every oxide surface investigated in this study, e.g., a glass plate (Fig. 1(a)), an STO crystal (Fig. 1(b)), and a complex iridium oxide ($Sr_2IrO_4$) thin film (Fig. 1(c)). This observation of [111]-orientation in PLD-grown Au films is consistent with a previous report of Ref. [13].

In contrast, for thermally evaporated Au films, there are four XRD peaks corresponding to the (111), (200), (220), and (311) planes of Au crystal, as shown in Fig. 2. It indicates that these Au films are polycrystalline with randomly orientated grains. All Au films that are deposited on other oxide surfaces through thermal evaporation exhibit similar random orientations in their XRD scans (data not shown).

### *3.2 Microscopic interfacial structures between Au films and oxides*

We investigated the interfaces between Au films and oxides using cross-sectional high-angle annular dark-field scanning transmission electron microscopy (HAADF-STEM). Figure 3(a) shows the interface between a PLD-grown Au film and an epitaxial $Sr_2IrO_4$ (SIO) thin film on an STO substrate. The bright region at the top Au film exhibits a single orientation with even distribution, which is consistent with the [111]-orientation of XRD shown in Fig. 1. Note that there is a flat narrow gap (< 1 nm) between Au and SIO, which is comparable to the interlayer spacing found in van der Waals materials.

In contrast, Figure 3(b) shows the interface between a thermally evaporated Au film and an STO substrate. The Au film shows a few grains with different orientations, which is also consistent with the XRD shown in Fig. 2. Note that there is an uneven gap of approximately > 3 nm between the Au film and the $SrTiO_3$ substrate. It is natural to ask how the reduced interfacial gaps observed in PLD-grown Au films will affect their adhesiveness, compared to thermally evaporated Au films.

*3.3 Simple adhesion test of Au films deposited on various surfaces*

PLD-grown Au films exhibit improved adhesion to oxide surfaces compared with thermally evaporated Au films. To qualitatively evaluate the adhesion strength of Au films deposited on different surfaces, we performed a simple adhesive tape test similar to the mechanical exfoliation procedures reported in Refs. [4, 14-16]. A piece of Scotch tape was applied to the Au film surface while carefully avoiding visible air pockets at the tape-film interface. A constant load of 1 g was then applied for 1 minute before the tape was removed.

As shown in Fig. 4, most of the Au films deposited by PLD on oxide thin films remained intact after the tape test, including the Au/SIO and Au/NIO samples. Furthermore, even when stronger adhesives such as Loctite Stycast were applied using a similar procedure, the PLD-grown Au films were not readily detached from the oxide surfaces. Similar behavior was consistently observed for numerous PLD-grown Au films deposited on various oxide materials, with most samples showing little or no visible damage after the peel-off test. These observations indicate that Au films deposited by PLD form substantially stronger adhesion to oxide thin films than is typically observed for conventionally deposited Au electrodes.

In contrast, PLD-grown Au films deposited on commercial glass substrates exhibited only partial resistance to the tape test. As shown in Fig. 4 (PLD Au/glass), large portions of the Au film were removed by the tape, although some regions remained attached to the substrate. This reduced adhesion may be related to the comparatively rougher and less well-defined surface of commercial glass compared with epitaxial oxide thin films, which could limit the formation of the highly ordered interfacial structure observed on oxide surfaces.

For comparison, Fig. 4 (Thermal Evaporation Au/glass) shows that thermally evaporated Au films were removed almost completely from the glass substrate during the tape test. These results suggest that PLD growth significantly enhances the adhesion of Au films, particularly on atomically flat oxide surfaces.

The tape test employed in this study provides only a qualitative assessment of adhesion and was not intended as a rigorous measurement of interfacial strength. Although the observed behavior consistently suggests enhanced adhesion of PLD-grown Au films on epitaxial oxide surfaces, quantitative evaluation of the adhesion energy and failure mechanisms will require dedicated measurements using established adhesion-testing techniques. Such a systematic study is beyond the scope of the present work and represents an important direction for future research.

### *3.4 Energetic PLD plasma and enhanced adhesion of Au films*

We propose that the high kinetic energy of the Au species within the PLD plume plays an important role in stabilizing the [111]-oriented Au films with enhancing their adhesion to oxide surfaces. Because PLD is an inherently non-equilibrium deposition process [11], its growth dynamics are complex and difficult to fully describe using existing theoretical models or simulations. Nevertheless, our experimental observations consistently demonstrate that Au films deposited from energetic laser-generated plumes develop a strong preferred orientation, whereas thermally evaporated Au films exhibit a more randomly oriented polycrystalline structure.

Previous studies have reported a systematic evolution from polycrystalline Au films deposited under less energetic PLD conditions to highly [111]-oriented Au films deposited under more energetic PLD conditions [13]. Our observation is consistent with them. The energetic Au species generated during laser ablation may promote surface diffusion and atomic rearrangement, facilitating the formation of the thermodynamically favorable Au (111) surface, which possesses the lowest surface energy among low-index Au crystal planes [17-20].

The highly textured microstructure may also contribute to the enhanced adhesion observed for PLD-grown Au films. In contrast to randomly oriented polycrystalline grains, a film consisting predominantly of [111]-oriented grains is expected to form a more uniform interface with an atomically flat oxide surface. Such an interface may permit stronger interfacial interactions and improved mechanical contact over a larger effective area [21, 22]. The atomically flat nanometer-scale gaps observed by STEM are consistent with the formation of a highly ordered metal-oxide interface. However, the present results do not establish whether this structural uniformity directly causes the observed improvement in adhesion.

In addition to promoting crystallographic texture, the energetic PLD plume may contribute to adhesion enhancement through plasma-related surface modification effects. Ion bombardment associated with energetic deposition processes has previously been reported to improve the adhesion of Au films [4]. Although the STEM images presented in Fig. 3 do not reveal any noticeable incorporation of Au atoms into the underlying oxide layers, the absence of observable interdiffusion does not exclude the possibility that plasma exposure during the initial stages of growth modifies the oxide surface. Similar plasma treatments are known to alter surface morphology and increase surface reactivity, thereby improving thin-film adhesion even in the absence of additional interfacial layers [6].

At present, the relative contributions of crystallographic texturing, interfacial ordering, and plasma-induced surface modification cannot be determined quantitatively. We therefore regard the mechanisms discussed above as plausible explanations that are consistent with the structural and adhesion data presented in this work. Further investigations using quantitative adhesion measurements together with detailed interfacial characterization will be necessary to identify the dominant mechanisms responsible for the enhanced adhesion of PLD-grown Au films on oxide surfaces.

**4. Conclusions**

Overall, this work demonstrates that pulsed laser deposition produces Au electrodes with a strong [111] preferred orientation, atomically flat metal-oxide interfaces, and enhanced adhesion to epitaxial oxide surfaces. These distinctive structural characteristics, which arise from the energetic nature of the PLD process, may provide important advantages for the integration of Au electrodes in oxide-based electronic devices. The ability to grow both epitaxial oxide films and high-quality Au electrodes within a single deposition platform offers a promising route toward simplified device fabrication and improved interface engineering. Future studies employing quantitative adhesion measurements and extending the approach to semiconductor substrates will further elucidate the broader applicability of this growth method.

**Declaration of Competing Interest**

The authors declare that they have no known competing financial interests or personal relationships that could have appeared to influence the work reported in this paper.

**Data availability**

Data will be made available on request.

**Acknowledgments**

We thank Jungho Kim and Pegor Aynajian for confirming to us that our additional PLD-grown Au films are strongly bonded to oxide thin-film surfaces. We acknowledge the support of the National Science Foundation Grants DMR-2104296, DMR-2426874 (R.L., S.S., T.T., A.S.) and DMR-1847964 (M.Z., C.C., J.H.) for sample synthesis and characterization.

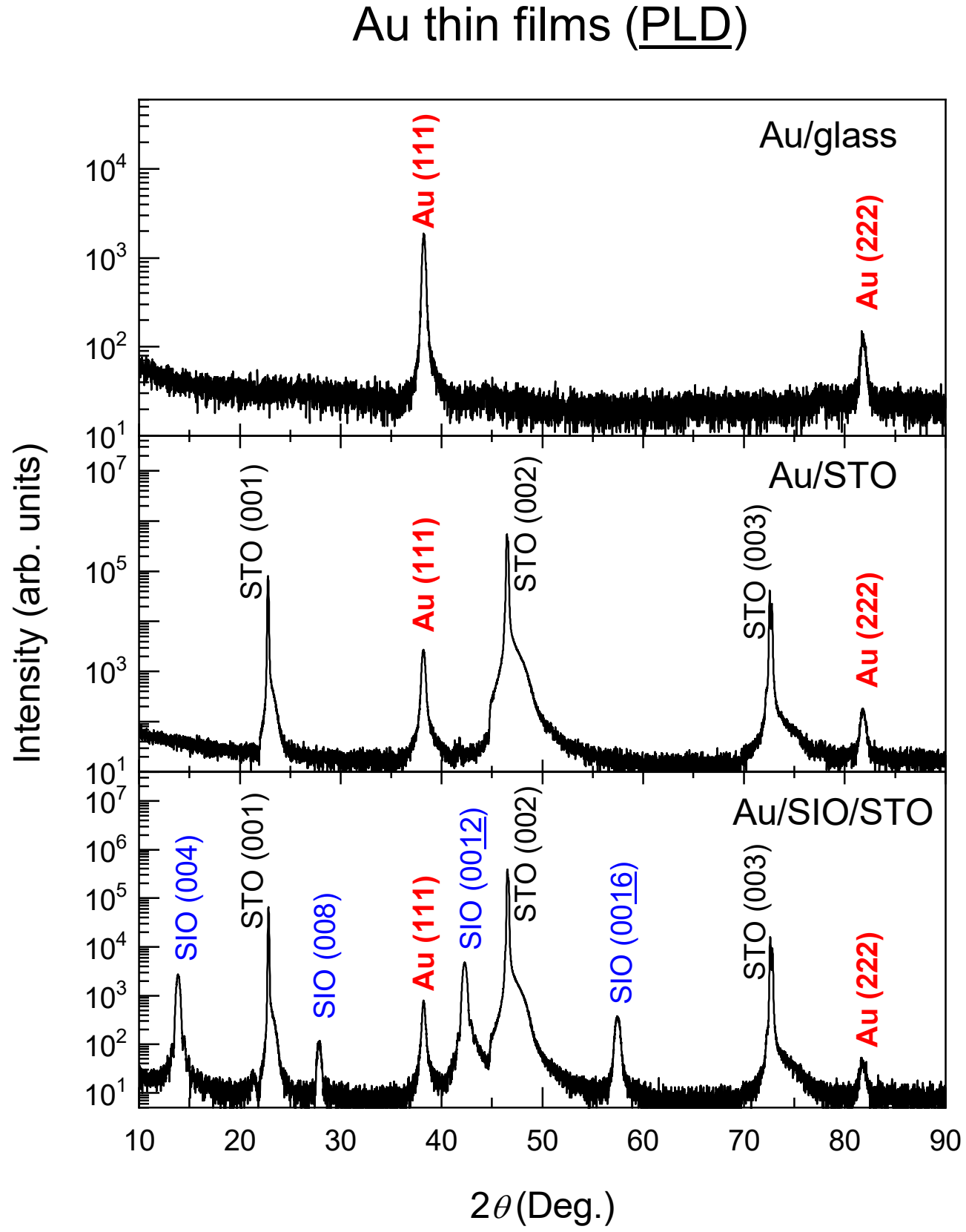

**FIG. 1.** XRD $\theta$-2$\theta$ scans of PLD-grown Au films on **(a)** a glass substrate, **(b)** a single crystal $SrTiO_3$ (STO) substrate, and **(c)** an epitaxial $Sr_2IrO_4$ (SIO) thin film on an STO substrate. All PLD-grown Au films demonstrate a preferred [111]-orientation along the surface normal direction.

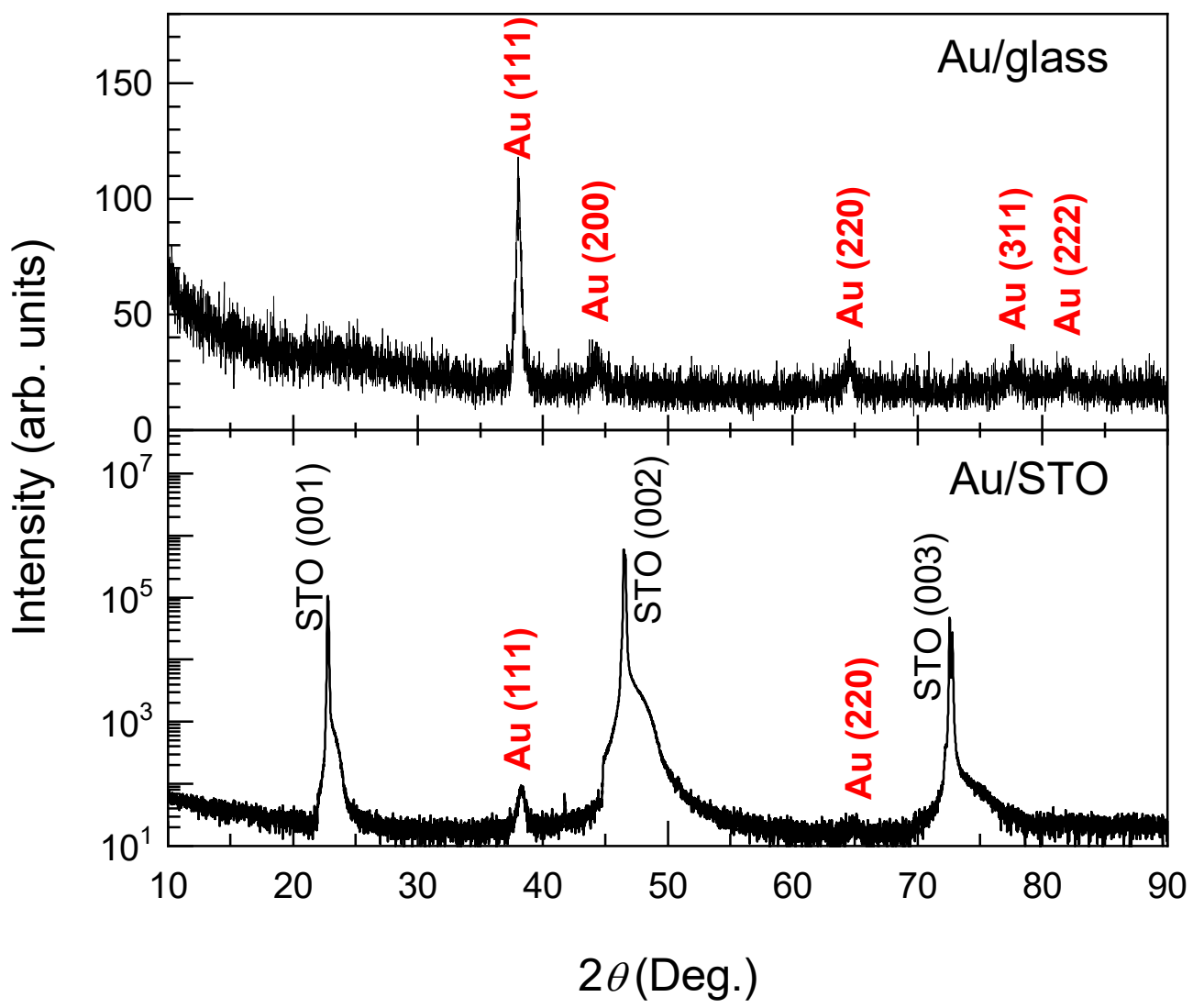

**FIG. 2.** XRD $\theta$-$2\theta$ scans of thermally evaporated Au films **a)** a glass substrate and **(b)** a single crystal STO substrate. In contrast to the PLD-grown samples shown in Fig. 1, these Au films are polycrystalline with random orientation.

**(a)** PLD

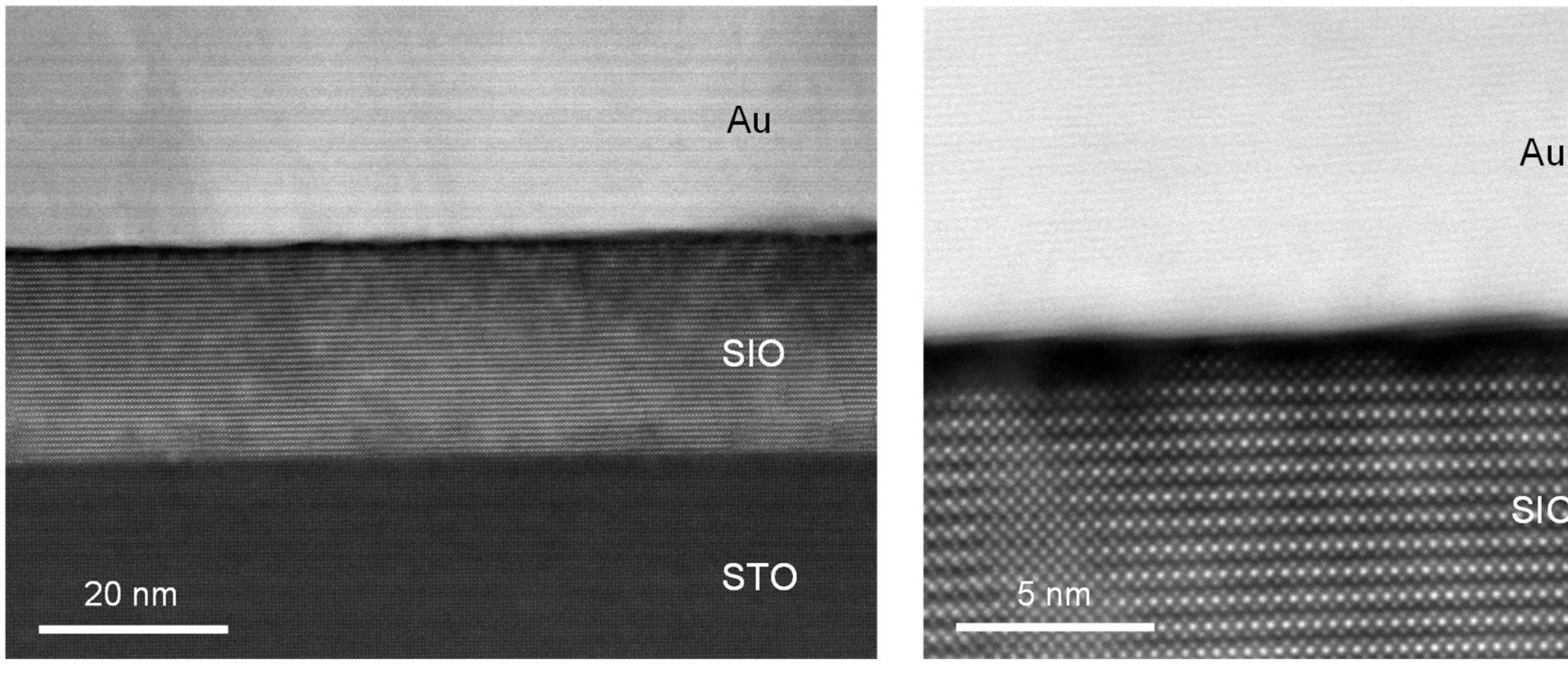

**(b)** Thermal evaporation

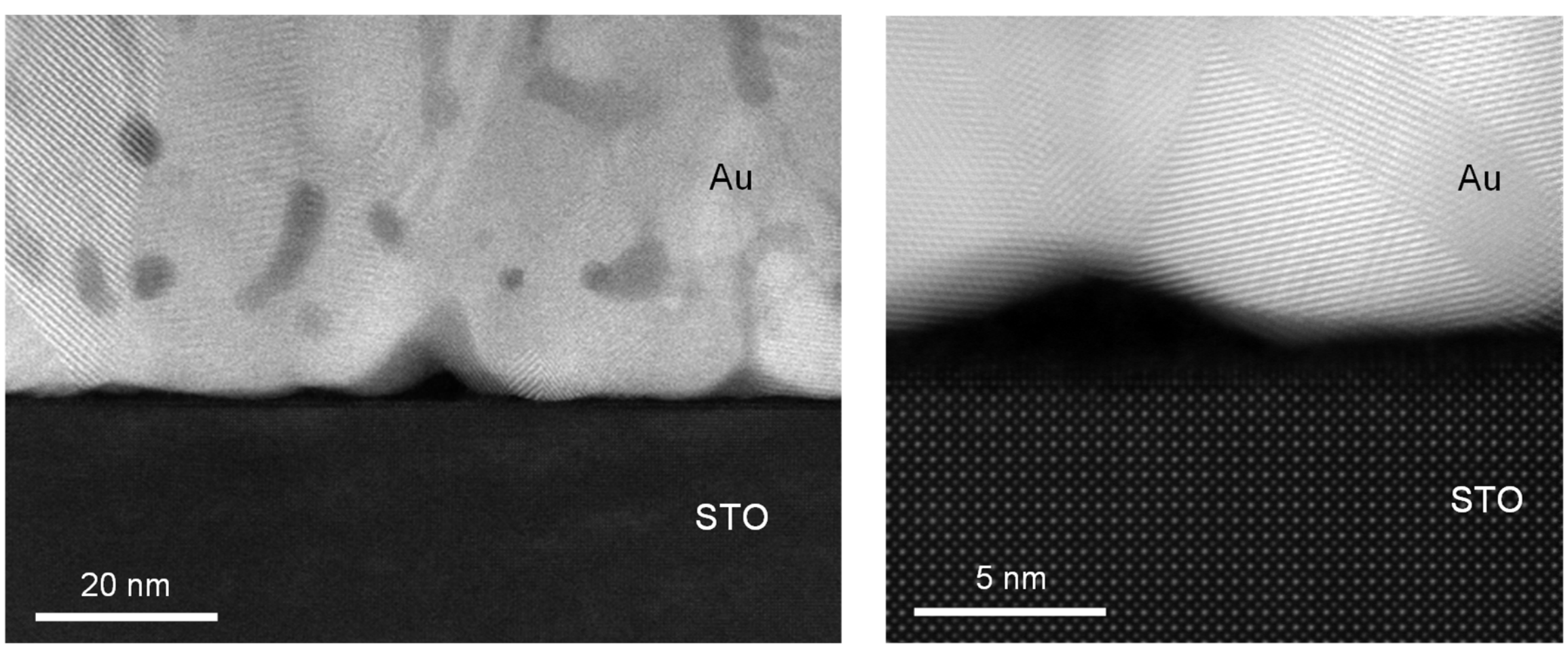

**FIG. 3.** Cross-sectional STEM images of **(a)** a PLD-grown Au film on SIO/STO and **(b)** a thermally evaporated Au film on STO. The left images show large areas with 20-nm scale bars. The right images are high-resolution scans near the interfaces between the Au films and oxides with 5-nm scale bars.

PLD Thermal evaporation

Before mechanical exfoliation

Au/SIO Au/NIO Au/glass Au/glass

After mechanical exfoliation

**FIG. 4.** Photographs of Au films on various substrates and surfaces before and after mechanical exfoliation, i.e., tape test. PLD-grown Au films on SIO and $Nd_2Ir_2O_7$ (NIO) thin films are not peeled off at all. Besides, both PLD-grown and thermally evaporated Au films on glass substrates are easily removed by tape presumably due to less adhesive interface between Au films and glass. The dimension of these samples is a few mm.